\documentclass[journal]{IEEEtran}

\ifCLASSINFOpdf

\else

\fi

\usepackage{amsmath}
\usepackage{graphicx}
\usepackage{amssymb}
\usepackage{amsthm}
\usepackage{cite}
\usepackage{mathtools}
\usepackage{bbold}
\usepackage{steinmetz}
\usepackage{stmaryrd}
\usepackage{bbm, dsfont}
\usepackage{verbatim}
\usepackage[latin1]{inputenc}
\usepackage{tikz}
\usetikzlibrary{shapes,arrows}
\usepackage{enumerate}
\usepackage{color}
\usepackage[caption=false,font=footnotesize]{subfig}
\usepackage{pgfplots}
\usetikzlibrary{calc}
\usepackage{mathrsfs}

\usepackage{steinmetz}
\usepackage[outdir=./]{epstopdf}

\usepackage{lipsum}
\usepackage[variablett]{lmodern}
\usepackage{amsbsy}
\usepackage{bm}
\usetikzlibrary{positioning}
\usetikzlibrary{shapes,arrows}
\usepackage{multirow}

\pgfplotsset{
	every tick label/.append style={scale=1},
	every axis/.append style={
	}
}

\pgfplotsset{
	grid style = {
		dash pattern = on 0.05mm off 1mm,
		line cap = round,
		black,
		line width = 0.5pt
	}
}
\usetikzlibrary{shapes.multipart}

\newcommand{\expe}{\mathrm E}  
\newcommand{\lefto}{\mathopen{}\left}    
\newcommand{\di}{\mathop{}\!\mathrm{d}}            
\begin{document}

\title{How to Increase the Achievable Information Rate by Per-Channel Dispersion Compensation}

\author{
	Kamran~Keykhosravi,~\IEEEmembership{Student~Member,~IEEE,}
	Marco~Secondini,~\IEEEmembership{Senior~Member,~IEEE,}
	Giuseppe~Durisi,~\IEEEmembership{Senior~Member,~IEEE,}
		and~Erik~Agrell,~\IEEEmembership{Fellow,~IEEE.}

	\thanks{ This work 
		was supported by the Swedish Research Council (VR) under Grants 2013-5271 and 2017-03702, and by the Ericsson Research Foundation. }
	\thanks{K. Keykhosravi, G. Durisi, and E. Agrell are with the Department of Electrical Engineering,
		Chalmers University of Technology, 41296 Gothenburg, Sweden (e-mail: kamrank@chalmers.se). 
		}
	\thanks{M. Secondini is with the TeCIP Institute, Scuola Superiore Sant'Anna, 56124 Pisa, Italy.
	}	
}

\maketitle

\begin{abstract}
	 Deploying periodic inline chromatic dispersion  compensation enables reducing the  complexity of the digital back propagation (DBP) algorithm.
	However, compared with nondispersion-managed (NDM) links,  dispersion-managed (DM) ones suffer a stronger cross-phase modulation (XPM).
	Utilizing per-channel dispersion-managed (CDM) links (e.g., using fiber Bragg grating) allows for a complexity reduction of DBP,  while abating XPM compared to DM links. 
	In this paper, we show for the first time that CDM links enable also a more effective XPM compensation compared to NDM ones, allowing a higher achievable information rate (AIR). This is explained by resorting to the frequency-resolved logarithmic perturbation model and showing that per-channel dispersion compensation increases the frequency correlation of the distortions induced by XPM over the channel bandwidth, making them more similar to a conventional phase noise. 
	 We compare the performance (in terms of the AIR) of  a DM, an NDM, and a CDM link, considering two types of mismatched receivers: one neglects the XPM phase distortion and the other compensates for it. 
	With the former, the CDM link is  inferior to the NDM one due to an increased in-band signal--noise interaction.
	However, with the latter, a higher AIR is obtained with the CDM link than with the NDM one owing to a higher XPM frequency correlation.
	 The DM link has the lowest AIR for both receivers because of a stronger XPM. 
\end{abstract}

\begin{IEEEkeywords}
Achievable information rate, fiber Bragg grating, optical communication, per-channel dispersion compensation,  XPM mitigation.
\end{IEEEkeywords}

\IEEEpeerreviewmaketitle

\section{Introduction}

\IEEEPARstart{T}{ransmission} over   long-haul fiber-optic systems is predominantly impaired by chromatic dispersion (CD), Kerr nonlinearity, and amplified spontaneous emission (ASE) noise \cite{Kessiambre_2010_jlt}.
Two general approaches for compensating CD are inline dispersion compensation (DC) and electronic DC. 
Systems in the former category mitigate the effects of CD via passive optical components installed before each amplifier. Depending on the components' dispersion profile, the effects of CD can be either removed locally within each wavelength-division-multiplexed (WDM) channel (e.g., via fiber Bragg grating (FBG)) or can be compensated for throughout the entire spectrum (via dispersion-compensating fibers (DCFs)). We refer to these two systems as per-channel dispersion-managed (CDM) and dispersion-managed (DM) links, respectively.
Systems with electronic DC, which are also referred to as nondispersion-managed (NDM) links, make use of digital signal processors (DSPs) to counter CD. 
This paper provides a  comparison among CDM, DM, and NDM links.

Over the last decade, DSPs have become a key element in  long-haul coherent optical systems. As CD can be effectively compensated for via DSPs, inline DC is not deployed in modern coherent systems since \textit{i)} it is not cost efficient, and \textit{ii)} it is believed to be detrimental to the system's  performance (see for example \cite[Sec.~XI-C]{Kessiambre2008capacity}). Nonetheless, studying inline DC methods is still relevant since  \textit{i)} they are used in systems where new coherent transmissions coexist with  legacy direct-detection ones, \textit{ii)} they reduce the channel memory and consequently allow for a complexity reduction of the digital back propagation (DBP) algorithm \cite{zhu2011folded,du2014channelized}, and \textit{iii)} they mitigate the effects of laser phase noise by reducing the equalization-enhanced phase noise \cite{Kcolavolpe2011impact}. In this paper, we show for the first time that  CDM  can also improve the  performance of the fiber optical systems. This might renew the interest in this technology for the development of new greenfield networks.

A number of studies have compared  inline and electronic DC systems.  In \cite{Kalfiad2009comparison} a polarization-multiplexed  return-to-zero differential quadrature phase-shift keying  signaling was considered and the bit-error rate was measured experimentally. In the absence of differential group delay, comparable results were reported for NDM and DM links.  
Several studies have shown that unlike NDM links, with inline dispersion-compensated systems, the complexity  of DBP can be significantly reduced via deploying folded DBP \cite{zhu2011folded,zhu2012nonlinearity,xia2014multi,du2014channelized}.   
 In \cite{du2014channelized}, the performance,  in terms of received signal-to-noise ratio (SNR), of a CDM link and an NDM link were compared via numerical simulations. 
For the  polarization-multiplexed  quadrature phase-shift keying modulation format, by deploying folded DBP,  the authors show that the CDM link can reach the same SNR as the NDM link with a much less complex receiver. 

In \cite{marco_arxive},  the frequency-resolved logarithmic perturbation model in \cite{Ksecondini2013achievable} was used to study  XPM coherence for distributed and NDM-lumped amplified systems. Furthermore, AIRs were calculated using a particle approach for NDM and DM links with phase and polarization noise compensation. Higher AIRs were obtained with the NDM link than with the DM one. 
This can also be seen with the setup in \cite{Keriksson2016impact}, where AIRs were calculated for DM and NDM links with polarization-multiplex  quadrature amplitude modulation for multiple auxiliary channels.	
In \cite{irukulapati2018improved} improved AIRs were obtained using an auxiliary backward channel for CDM links.

This paper goes beyond the existing literature by providing a comparison between the performance of all the three links ( CDM,   DM, and  NDM links) in terms of the achievable information rate (AIR). We assume that the intra-channel signal--signal distortion is compensated for via DBP. In this case, cross-phase modulation (XPM) \cite[Ch.~7]{Kagrawal_2007_nfo} becomes the predominant  nonlinear impairment \cite{bononi1998impulse,Kessiambre_2010_jlt,Kmecozzi2012nonlinear,Kdar_2013_oe}.  
The first part of this paper is devoted to  studying the properties  of XPM. We adopt the channel model developed in \cite{Ksecondini2013achievable} for  NDM and  DM links and extend it to the CDM case. By doing so, we compare the variance and the autocorrelation of the XPM distortion in the three links. We show that the DM link suffers from a much stronger XPM  compared with the NDM and  CDM links, for which  XPM  has the same variance. Furthermore, we show, for the first time, that with the CDM link,  XPM  has a damped periodic temporal correlation and also has a higher frequency correlation compared with the NDM link.

In the second part of the paper, we assess the performance of the three links   by evaluating  AIRs.
Calculating  the AIR is a common approach to obtain lower bounds on the capacity of the fiber-optic channel, whose  exact capacity is unknown \cite{Ksecondini2017scope}. In order to calculate the AIR, one needs to select an input distribution and an auxiliary channel law. The AIR then determines the rate achievable on the actual fiber channel via the mismatched detector optimized for the auxiliary channel \cite{Kmerhav1994information,Ksecondini2017scope,Karnold2006simulation}. In this paper, we fix the input distribution to be zero-mean Gaussian and  consider two different  types of auxiliary channels. One is an additive white Gaussian noise (AWGN) model and the other is a phase-noise model. While the former does not consider the XPM phase noise, the latter does so by modeling XPM as an autoregressive (AR) phase-noise process of order one. The AIR calculated based on these two models can be translated  into the rates achievable by two mismatched receivers, where only the second one compensates for XPM. Our results indicate that mitigating XPM by exploiting its temporal correlation improves the AIR significantly, which is in agreement with previous studies \cite{Kdar2017nonlinear,Kdar2013improved,Ksecondini2012analytical,Kdar_2014_ol,Kdar2015inter}. This also highlights  the fact that the Gaussian-noise models (see for example \cite{Ksplett_1993_ecoc,Kpoggiolini_2011_ptl2,Kjohannisson_2013_jlt}) do not accurately represent nonlinear distortions, a point made previously in \cite{Kdar_2013_oe}.

 With both receivers, the DM link has an inferior AIR compared to the NDM and  CDM links due to a stronger XPM.   
 Furthermore, we found out that the outcome of the performance comparison between the NDM and  CDM links depends on the type of  receiver (or equivalently, type of  auxiliary channel). 
 With the receiver optimized for the AWGN channel, the CDM link is inferior to the NDM one as it induces a stronger in-band signal--noise interaction.  
 On the contrary, with the receiver that compensates for  XPM, the CDM link prevails due to a higher XPM spectral  coherence. 
Previous works often  optimize either the receiving algorithm (e.g.,\cite{maher2015modulation,keykhosravi2018demodulation}) or the transmission line \cite{Kip_2008_jlt,du2014channelized}. Our results indicate that optimizing the transmission line in conjugation with the receiver leads to an additional performance gain.
Furthermore, motivated by the shape of the XPM autocorrelation function calculated in the first part of the paper for the CDM link, we study a third auxiliary channel, in which the XPM phase distortion is modeled as an AR  process of order higher than one.   This results in a further improvement of the AIR for the CDM\footnote{We observed that for the NDM and  DM links, this auxiliary channel does not improve the AIRs compared with the AR model of order one.} link. To the best of our knowledge, this is the first time that such an auxiliary channel is studied in optical literature.

The remainder of this paper is structured as follows: 
In Section~\ref{sec:Channel_model}, we introduce the channel model that has been proposed in \cite{Ksecondini2013achievable} for NDM and DM links. Furthermore, we extend this model to cover also  CDM links. 
In Section~\ref{sec:XPM_TF_coh}, an expression for the XPM time and frequency correlation is presented and numerically evaluated for the NDM,  DM, and  CDM links. 
The performance of these three links is assessed in Section~\ref{sec:XPM_mit_AIR_cal} by evaluating AIRs.
Finally, Section~\ref{sec:conclusion} concludes the paper.

\section{Modeling XPM Distortion} \label{sec:Channel_model}

{In this section, we investigate the channel model proposed in \cite{Ksecondini2013achievable}. We focus on the  effects of XPM and neglect the ASE noise. The results of this section are used in Section~\ref{sec:XPM_TF_coh} to analyze the properties of XPM. We deploy this analysis to explain the simulation results in Section~\ref{sec:XPM_mit_AIR_cal}, where  WDM systems are simulated via the split-step Fourier method and  the ASE noise is included.}

\begin{figure}[!t]
	\centering

	\includegraphics[]{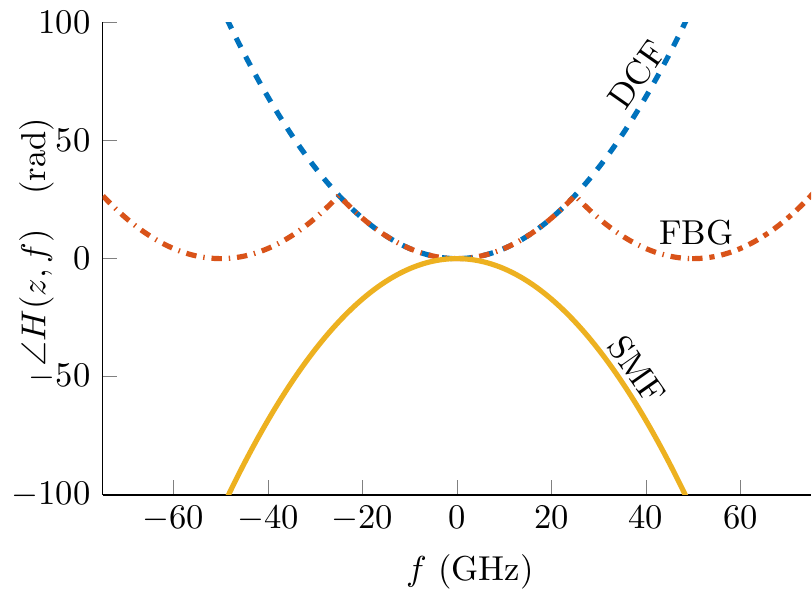}

	\caption{Phase of the CD transfer function for a single span of $100$ km of SMF, and for the corresponding DCF and FBG as DC components.}
	\label{CD_map}
\end{figure}

Denote by $u(z,t)$ the  complex envelope of the signal transmitted over the channel of interest (COI) of a WDM system at time $t$ and location $z$.  Moreover, let $w(z,t)$ indicate the aggregation of all interfering signals. The propagation of  $u(z,t)$ through a single-polarization fiber-optic system is governed by the equation \cite{Kmitra2001nonlinear,Ksecondini2013achievable,Ksecondini2012analytical}
\begin{IEEEeqnarray}{c}\label{eq_nlse2}
	\frac{\partial u}{\partial z}=j\frac{\tilde{\beta}_2(z)}{2}\frac{\partial^2u}{\partial t^2}-j\gamma\left(a_u|u|^2+2a_w|w|^2\right)u.
\end{IEEEeqnarray}
Here, the coefficients $a_u(z)$ and $a_w(z)$ determine the power of the signals $u$ and $w$, respectively, at location $z$ normalized by the input power and account for the attenuation or amplification effects throughout the propagation. Specifically,  $a_u(z)=a_w(z)=\exp(-\alpha(z \mod L_s))$, where $L_s$ denotes the span length and $\alpha$ is the attenuation constant of the standard single-mode fiber (SMF). The constant $\gamma$ in \eqref{eq_nlse2} is the nonlinear coefficient and $\tilde{\beta}_2(z)$ denotes the CD parameter at location $z$. For SMF $\tilde{\beta}_2(z)=\beta_2$, where $\beta_2$ is the fiber's CD parameter. When a FBG or a  DCF is installed at the end of the $k$th span, we have that $\tilde{\beta}_2(z)=-L_s\beta_2\delta(z-kL_s)$, where  $\delta(\cdot)$ is the Dirac delta function.
We shall neglect the attenuation and the nonlinearity of FBG and DCF.
 We  assume that the  intra-channel signal--signal distortions are compensated for perfectly by applying DBP to the COI at the receiver.
 By replacing the terms $|u|^2$ and $|w|^2$  in \eqref{eq_nlse2} with their linearly propagated counterparts and by exercising the first-order logarithmic-perturbative method, the approximate channel model 
\begin{IEEEeqnarray}{c}
	\tilde u(L,t)\approx \int_{-\infty}^{\infty}U(f)e^{-j\theta(f,t)} e^{j2\pi ft}\di f
\end{IEEEeqnarray}
is obtained \cite{Ksecondini2012analytical,Ksecondini2013achievable}.
Here, $L$ is the system length and $\tilde u(L,t)$ indicates the received signal after DBP.
$U(f)$ represents the Fourier transform of $u(0,t)$.  The XPM term $\theta$ is
   \begin{align}\label{eq_theta}
   	\theta(f,t)=2\iint\limits_{\mathbb{R}^2}K_w\lefto(f,\mu,\nu\right)W(\mu)W^*(\nu)e^{j2\pi(\mu-\nu)t}\di \mu\di \nu
   \end{align}
   where $W(f)$ is the Fourier transform of $w(0,t)$. Also, 
   \begin{align}\label{eq_kw}
   	   	&K_w\lefto(f,\mu,\nu\right)= \nonumber\\
   	   	&\ \  \gamma\int_{0}^{L}a_w(z)H(z,\mu)H^*(z,\nu)H(z,f)H^*(z,\mu-\nu+f)\di z
   \end{align}
where
\begin{IEEEeqnarray}{c}\label{eq_hzf}
	H(z,f)=\exp\lefto(-j2\pi^2f^2\int_{0}^{z}\kappa(\zeta,f)\di\zeta\right)
\end{IEEEeqnarray}
indicates the CD transfer function from the transmitter to distance $z$. 
Here, $\kappa(\zeta,f)$ captures the changes in the dispersion profile in both frequency and space.
 With SMF, $\kappa(\zeta,f)$ is constant and  $\kappa(\zeta,f)=\beta_2$. We consider two other components that affect $H(z,f)$, namely, DCF and FBG.  A DCF installed at the end of the $k$th span can be modeled by setting $\kappa(\zeta,f)=-L_s\beta_2\delta(\zeta-kL_s)$ in \eqref{eq_hzf}. The FBG at the end of the $k$th span can be modeled by setting
\begin{IEEEeqnarray}{c}\label{eq_FBG_beta}
	\kappa(\zeta,f)=-\left({\tilde{f}}/{f}\right)^2L_s\beta_2\delta(\zeta-kL_s)
\end{IEEEeqnarray}
where
\begin{IEEEeqnarray}{c}\label{eq_ftild}
	\tilde{f}=\min_{m\in\mathbb{Z}}\left|f-mB\right|
\end{IEEEeqnarray}
and $B$ is the channel bandwidth. Fig.~\ref{CD_map} depicts the phase of the transfer function $H(z,f)$ for a $100$-km standard SMF and also for the corresponding DCF and FBG components in a $50$-GHz WDM grid.  

For  the CDM link with $N_s$ spans, by substituting \eqref{eq_hzf}--\eqref{eq_ftild} into \eqref{eq_kw}, we obtain  after some standard algebraic steps
\begin{IEEEeqnarray}{rl}\label{eq_kw_fbg}
	K_w&\lefto(f,\mu,\nu\right)=  \gamma\frac{\exp\lefto\{\left(-\alpha +jg\lefto(f,\mu,\nu\right)\right)L_s\right\}-1}{-\alpha+jg\lefto(f,\mu,\nu\right)}\nonumber\\
	&\cdot\sum\limits_{n=0}^{N_s-1}\exp\lefto(jnL_s\lefto(g\lefto(f,\mu,\nu\right)-g\lefto(\tilde f,\tilde \mu,\tilde \nu\right)\right)\right).
\end{IEEEeqnarray}
Here, $\tilde{\mu}$ and $\tilde{\nu}$ are functions of $\mu$ and $\nu$ defined similarly as in \eqref{eq_ftild} and
\begin{IEEEeqnarray}{c}
	g\lefto(f,\mu,\nu\right)=4\pi^2\beta_2(\nu-f)(\nu-\mu).
\end{IEEEeqnarray}
With  DM links,   one needs to replace $g(\tilde f,\tilde \mu,\tilde \nu)$ by $g(f,\mu,\nu)$ in the summation in \eqref{eq_kw_fbg}, which simplifies to the constant $N_s$. With NDM links, $K_w\lefto(f,\mu,\nu\right)$ can be calculated by omitting the term $g\lefto(\tilde f,\tilde \mu,\tilde \nu\right)$ in \eqref{eq_kw_fbg}. 
For these two systems, the corresponding channel models are special cases of \cite[Eq.~(11)]{Ksecondini2013achievable}.

\section{XPM Time--Frequency Coherence  }\label{sec:XPM_TF_coh}
To characterize the coherence of the XPM distortion, we calculate its autocorrelation function as
\begin{IEEEeqnarray}{rCl}\label{eq_R}
	R_\theta(f_1,f_2,\tau,t)&=&
	\expe\lefto[\theta\lefto(t,f_1\right)\theta^*\lefto(t+\tau,f_2\right)\right]\nonumber\\
	&&-\expe\lefto[\theta\lefto(t,f_1\right)\right]\expe\lefto[\theta^*\lefto(t+\tau,f_2\right)\right].
\end{IEEEeqnarray}
Substituting \eqref{eq_theta} into \eqref{eq_R} we obtain a four-fold integral containing a forth-order moment of $W$. To proceed, similarly as in \cite{Ksecondini2013achievable}, we assume that $w$ is a stationary Gaussian process with power spectral density $S_w(f)=P_w/(2B_w)\mathrm{rect}((|f|-f_w)/B_w)$, where $f_w$ and $B_w$ represent the center frequency (for \mbox{$f>0$}) and the bandwidth of the interfering signal, respectively. Using Isserlis's theorem \cite[Eq.~(7-61)]{Kpapolis_PrSt} to decompose the fourth-order moment of $W$ into second-order moments and the equality $\expe[W(\mu)W^*(\nu)]=S_w(\mu)\delta(\mu-\nu)$, we obtain
 \begin{align}\label{eq_R1}
 	&R_\theta(f_1,f_2,\tau)=\nonumber\\
 	&\frac{P_w^2}{B_w^2}
 	\iint_{V^2}
 	 K_w\lefto(f_1,\mu,\nu\right)
 	 K_w^*\lefto(f_2,\mu,\nu\right)e^{-j2\pi(\mu-\nu
 		)\tau}\di \mu\di\nu.
 \end{align}
 Here, $V=T_{f_w}\cup T_{-f_w}$, where $T_f=[f-B_w/2 , f+B_w/2 ]$. Also, we have omitted the parameter $t$ on the right-hand-side of \eqref{eq_R1} as it is irrelevant to the calculation of the autocorrelation function because of stationarity.

\begin{table}[!t]
	\renewcommand{\arraystretch}{1.3}
	\caption{ System parameters used in the numerical examples.}
	\label{t2}
	\centering

	\caption{Correlation function (arbitrary unit)  of the XPM phase distortion $\expe[\theta(0,t)\theta^*(\Delta_f,t+\tau)]$ for  NDM,  DM, and  CDM links with three WDM channels. The cross-sections of the three countour plots at $\tau=0$ and $\Delta_f=0$ are compared in parts (d) and (e), respectively.  }
	\label{Fig_XPM_Corr_3ch}
\end{figure*}

To evaluate the XPM autocorrelation function,  we resort to numerical integration to calculate \eqref{eq_R1}. Furthermore, similar to \cite{Ksecondini2013achievable}, to reduce computational complexity, we approximate \eqref{eq_R1} by calculating the integration over   $T_{f_w}^2\cup T_{-f_w}^2$ instead of $V^2$ (we neglect the cross-products created by two different frequency bands $T_{f_w}$ and $T_{-f_w}$). 
We consider a multi-span fiber-optic system whose parameters are listed in Table~\ref{t2}. Here, $D=-2\pi c \beta_2/\lambda^2$, where $c$ is the speed of the light and $\lambda$ is the wavelength associated with the center frequency. We begin by studying three copropagating wavelengths, and then we analyze the results for five copropagating wavelengths. For both cases, the middle channel is selected as COI.

Fig.~\ref{Fig_XPM_Corr_3ch} depicts the  autocorrelation function in \eqref{eq_R1} for three  WDM channels. We fix $f_1=0$ and illustrate the  autocorrelation function $R_\theta(0,\Delta_f,\tau)$ via contour plots in Fig.~\ref{Fig_XPM_Corr_3ch}\,(a)--(c) (values are normalized). 
The temporal and spectral cross sections are depicted in  Fig.~\ref{Fig_XPM_Corr_3ch}\,(d) and (e), respectively; in both figures the three curves are normalized such that their overall maximum is one. 

 Fig.~\ref{Fig_XPM_Corr_3ch}\,(d) depicts $R_\theta(0,\Delta_f,0)$  for $|\Delta_f|\leq 25$ GHz. It can be seen that with the DM and  CDM links, the spectral correlation of  XPM is substantial across the bandwidth. On the other hand, when no inline DC is employed,  the correlation between the XPM frequency components decreases quickly with $\Delta_f$. 
 With the NDM link, due to CD, distinct signal frequency components  propagate through the fiber with differrent velocities, resulting in a time delay among them. Therefore, each  frequency component is corrupted by different realizations of interference caused by its neighboring channels.
 The larger the  gap between two frequencies, the greater the velocity divergence, and the weaker the  correlation between them. With the DM and  CDM links, the time delay  between the frequency components of the signal, caused by CD, is compensated for at the end of each span. Therefore, the signal experiences roughly the same interference across its spectral bandwidth. Hence, the frequency correlation is strong.
 
\begin{figure}[!t]
	\centering

	\begin{tikzpicture}[every text node part/.style={align=center}]

	\node at (11.9cm,-3.72cm){\includegraphics{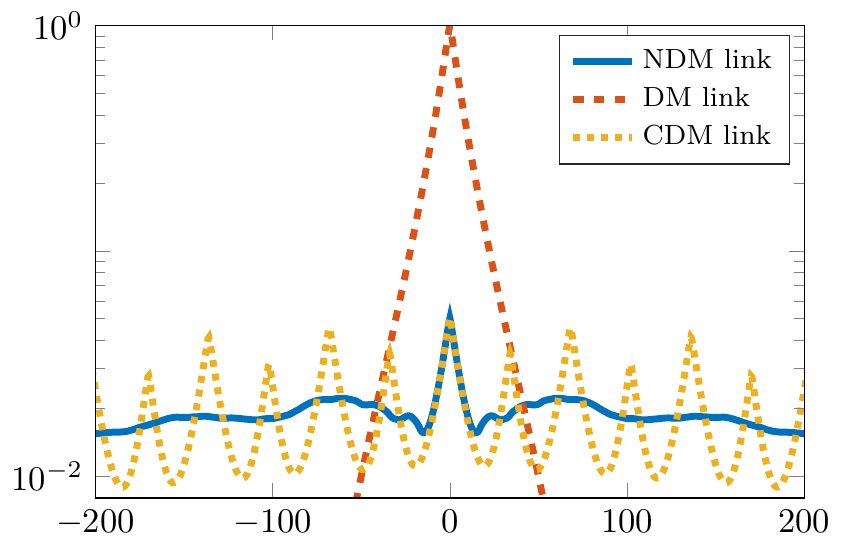}};

	\node at (12.2cm ,-6.7cm){\footnotesize Time separation $\tau$ (symbols)};
	
	\node[rotate=90] at (8cm ,-3.5cm){\footnotesize   Correlation};
	
	\node at (12.5cm ,-1cm){\footnotesize  Cross-sections  at $\Delta_f=0$ };

	\end{tikzpicture}
	\caption{Normalized temporal correlation function of the XPM phase distortion  link with five WDM channels.  }
	\label{Fig_XPM_Corr_5ch}
\end{figure}

\begin{figure*}[!t]
	\centering

	\begin{tikzpicture}[every text node part/.style={align=center}]

	\node at (2.5cm ,5cm){\includegraphics[scale=.7]{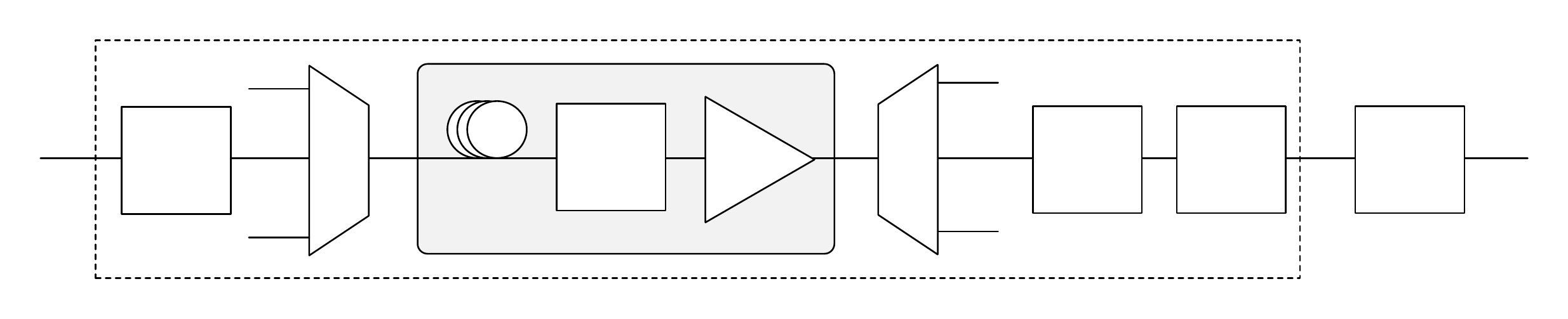}};
	
	\node at (-5.5cm ,5.2cm){ $x_l$};
	\node at (10.5cm ,5.2cm){ $\hat{x}_l$};
	\node at (8.6cm ,5.2cm){ $y_l$};
	\node at (-4.3cm ,5cm){ \footnotesize Mod.};
	\node at (7.5cm ,5cm){ \footnotesize MFS};
	\node at (5.9cm ,5cm){ \footnotesize DBP};
	\node at (9.5cm ,5cm){ \footnotesize Detector};
	\node[rotate=90] at (-2.5cm ,5cm){ \footnotesize MUX};
	\node[rotate=0] at (-.8cm ,5.84cm){ \footnotesize SMF};
	\node at (.5cm ,5cm){ \footnotesize DC};
	\node at (2.1cm ,5cm){ \footnotesize EDFA};
	\node[rotate=90] at (3.9cm ,5cm){ \footnotesize DEMUX};
	\node at (-3.3cm ,5.2cm){ \footnotesize Tx.2};
	\node at (-3.3cm ,6cm){ \footnotesize Tx.1};
	\node at (-3.3cm ,4.3cm){ \footnotesize Tx.3};
	\node at (4.6cm ,5.2cm){ \footnotesize Rx.2};
	\node at (4.6cm ,6.05cm){ \footnotesize Rx.1};
	\node at (4.6cm ,4.4cm){ \footnotesize Rx.3};
	\node at (3.35cm ,6cm){ \footnotesize $\times N_s$};
	\node at (1.7cm ,6.5cm){ \footnotesize Discrete-time channel};
	\end{tikzpicture}
	
	\caption{A schematic of the under studied WDM system model with three channels. Mod.: modulator; DC: dispersion compensator; MFS: matched filtering and sampling demodulator, EDFA: erbium-doped fiber amplifier. }
	\label{fig:System_model}
\end{figure*}
 
 As it is apparent from Figs.~\ref{Fig_XPM_Corr_3ch}\,(d) and (e), the XPM variance  $R_\theta(0,0,0)$ with the DM link is much larger compared to that with the NDM or  CDM links.
  With the DM  link, roughly no walk-off (i.e., the group-velocity difference between WDM channels) occurs between the interfering channels and the COI. Therefore, the XPM products aggregate coherently, resulting in an increased XPM variance.
  It can be seen from Fig.~\ref{Fig_XPM_Corr_3ch}\,(e) that  with the NDM and  DM links, the XPM temporal correlation  drops with $\tau$.
   With the CDM link, however,  the temporal XPM autocorrelation function behaves in a damped periodic fashion. The period  is roughly equal to the walk-off time between the COI and the two interfering channels across one span, that is, $T_p=D\Delta_\lambda L_s\approx 681$~ps ($\approx 34$ symbols), where $\Delta_\lambda$ is the WDM wavelength separation.
  Therefore, symbols that are $T_p$  apart, experience roughly the same set of interfering signals after each amplification, where the XPM distortion is at its strongest. 

Fig.~\ref{Fig_XPM_Corr_5ch} depicts the temporal XPM correlation for the WDM system described in Table~\ref{t2} with five channels. With the NDM and  DM links, a similar behavior as in Fig.~\ref{Fig_XPM_Corr_3ch}\,(e) can be observed. With the CDM link, the autocorrelation function is the sum of two damped periodic functions, one with a period of $T_p$ and the other with a period of $2T_p$. The former is brought about by the two  channels neighboring the COI and the latter by the two distant ones.

\section{XPM mitigation and  AIR calculation}\label{sec:XPM_mit_AIR_cal}

In this section, we evaluate and compare the AIR {(see for example \cite[Eq.\,(5)]{Ksecondini2017scope})} as a figure of merit for the three links described in Section~\ref{sec:XPM_TF_coh}. The discrete-time channel over which the AIR is calculated is illustrated in Fig.~\ref{fig:System_model}. 
To calculate the  AIR, we need to fix an input distribution and an auxiliary channel. Throughout the paper, we set the input distribution to be a zero-mean complex Gaussian. We  consider three auxiliary channels, which are specified in the following section.  The purpose of  the auxiliary channels is not only to calculate AIR but also to provide a guideline for designing better receivers. A typical approach to do so is to perform iterative soft-input soft-output detection and decoding, where the detector computes detection metrics based on the auxiliary channel model (see for example \cite{Kcolavolpe2005algorithms}).

\begin{figure*}[!t]
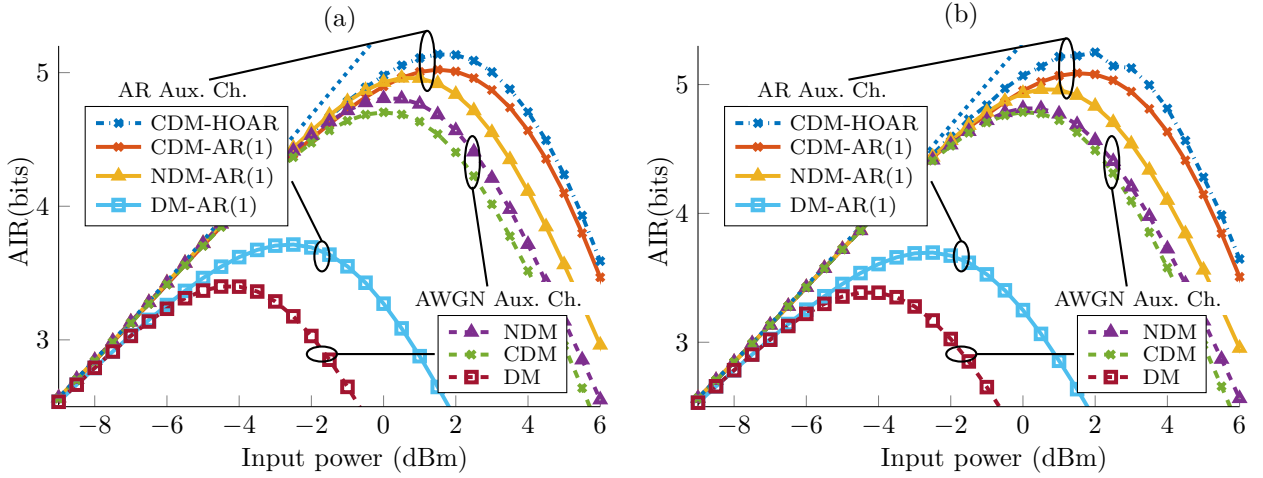

	\centering


	\caption{AIRs  for a 50-GHz WDM grid.  The ASE noise is injected (a) after each amplifier (b) at the transmitter. The capacity of the corresponding AWGN channel is  shown (dotted line) for comparison. }
	\label{AIR_50G}
\end{figure*}

\begin{figure*}[!t]
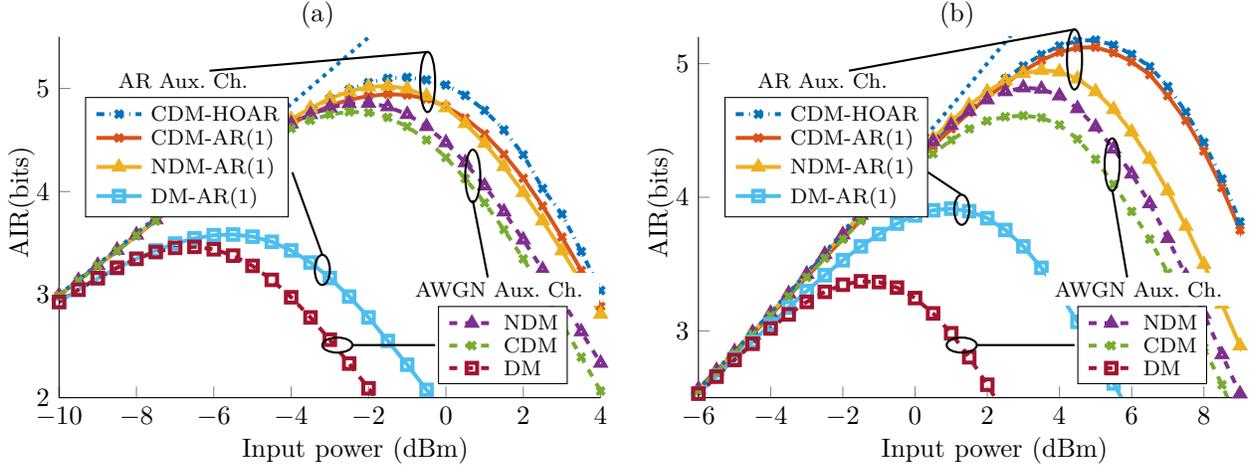

	\centering

	%
	\caption{AIRs  for a (a) 28-GHz and (b) 100-GHz WDM grid.  }
	\label{AIR_28_100}
\end{figure*}

\subsection{Auxiliary channel models}\label{sec:sub:aux_ch_model}

Similarly as in \cite{marsella2015simple}, we consider the following frequency-flat discrete-time input--output relation to serve as an auxiliary channel in calculating the AIR:
\begin{equation}\label{aux.ch}
y_l=h_0x_le^{j\theta_l}+n_l
\end{equation}
where $l$ is the time index, $h_0\in \mathbb{R}$ is the channel coefficient, \mbox{$\theta_l\in \mathbb{R}$}  accounts for the XPM phase distortion,  ${n}_l\in \mathbb{C}$ models a complex additive noise, and $x_l$ and $y_l$ denote  the complex channel input and output, respectively. We assume that  $n_l$ follows an independent and identically distributed  circularly-symmetric  Gaussian distribution with variance  $\sigma^2_n$.
In this paper, we study the following three  auxiliary channels based on the distribution imposed on~$\theta_l$.

\begin{enumerate}
	\item \emph{AWGN model:} this channel is simply obtained by neglecting the XPM phase distortion and setting $\theta_l=0, \ \ \forall l$ in \eqref{aux.ch}. 
	\item \emph{Autoregressive model of order 1 (AR(1)):} The random process $\{\theta_l$\} is modeled as 
	\begin{equation}\label{dd}
		\theta_l=\theta_{l-1}+z_l  \ \ \mod 2\pi
	\end{equation}
	where $z_l$ is an independent and identically distributed real  Gaussian process with variance $\sigma^2_z$. We note that \eqref{dd} corresponds to  a discrete-time Wiener process.  
	\item \emph{Higher-order autoregressive model (HOAR):} The random process $\theta_l$ is modeled as 
		\begin{equation}
		\theta_l=\alpha\theta_{l-1}+(1-\alpha)\theta_{l-l_0}+z_l \ \ \mod 2\pi
		\end{equation}
where $0\leq\alpha\leq 1$, $l_0>1$, and $z_l$ is distributed similarly as in AR(1). This model is motivated by the temporal correlation  of the CDM link in Fig.~\ref{Fig_XPM_Corr_3ch}\,(e) in order to create a damped periodic autocorrelation function.
\end{enumerate}
The AIR calculated based on the  AWGN auxiliary model can be obtained by a receiver that neglects the XPM phase distortion. Here, the AIR is calculated  using \cite[Eq.~(6)]{Ksecondini2017scope}. On the contrary, the receivers optimized for the AR(1)  and  HOAR models compensate for  XPM. In this case, the AIR  is evaluated using the particle approach proposed in \cite{dauwels2008computation}, which was applied to NDM and DM fiber-optic links with phase and polarization noise in  \cite{marco_arxive}.

\subsection{Numerical example}\label{sec:sub:numerical_ex}
We evaluate   the AIR for the lumped-amplified system with parameters  in Table~\ref{t2}. First, we show the results for three and then for five WDM channels. A total number of $10^5$ symbols are transmitted, out of which the first $2000$ are used to optimize the parameters ($h_0, \sigma_n, \sigma_z, \alpha, l_0$) of the auxiliary channels. 
The parameter $\sigma_n$ is estimated, as $\sigma_n=\max_{\sigma_n}\sum_i \log P(|y_i|^2\,|\,x_i,\sigma_n,h_0)$, where the likelihood $P(|y_i|^2\,|\,x_i,\sigma_n,h_0)$ is calculated based on a non-central chi distribution and $h_0$ is estimated as follows: $h_0^2=\sum_{i} (|y_i|^2-\sigma^2_n)/{\sum_{i}|x_i|^2}$.
The rest of the parameters ($ \sigma_z, \alpha, l_0$) are optimized using a genetic optimization algorithm  that attempts to maximize the AIR. After optimizing the parameters, the AIR estimation is  performed based on the remaining 98,000 symbols. 
 Symbols are drawn from a complex Gaussian distribution and modulation is performed via sinc pulses. The optical fiber is simulated by  means of the split-step Fourier method\footnote{ In order to ensure the accuracy of the split-step Fourier simulations, the number of steps and sampling rate are selected such that  increasing them results in negligible impact on the output.} \cite[Ch.~2]{Kagrawal_2007_nfo}.

Fig.~\ref{AIR_50G}\,(a) illustrates the AIR for the three links with three 50-GHz WDM channels.
 The profound influence of XPM mitigation on the AIR can be observed by comparing the rates achieved via the AWGN auxiliary channel model with those obtained by the AR  (AR(1) and HOAR)\footnote{We observed no improvement by considering the HOAR auxiliary  channel instead of the AR(1) for the NDM  and  DM links. } models. 
 In all cases, the AIR is substantially lower with the DM link compared to the NDM and  CDM ones. 
This is due to the periodic compensation of the walk-off between channels in the DM link, which increases the variance of the XPM distortion, as shown by the autocorrelation functions $	R_\theta(0,\Delta f,0)$ in Fig.~\ref{Fig_XPM_Corr_3ch}\,(d) at $\Delta_f=0$. 

 Fig.~\ref{AIR_50G}\,(a) also shows that, with the AWGN auxiliary channel, the CDM link is inferior to the NDM one, while with the AR(1) model, the opposite behavior is observed. We focus first on the AWGN auxiliary channel. As it is evident from Fig.~\ref{Fig_XPM_Corr_3ch}\,(d), the variance of the XPM distortion at the central frequency of COI $	R_\theta(0,0,0)$ is roughly equal for both the NDM and the CDM link.\footnote{Based on our numerical evaluation (not included in this paper), the XPM variance is roughly the same for both the NDM and  CDM links across the COI spectrum (not only at the central frequency).} Therefore, the XPM effects are not responsible for the difference between the AIRs. This gap can be explained through the nonlinear phase noise (NLPN) induced by  self-phase modulation (SPM) \cite[Fig.~27]{Kessiambre_2010_jlt}, that is, the signal--noise interaction within the bandwidth of the COI.
 Since with  the CDM link the dispersion is compensated for within each WDM channel, the intrachannel nonlinear products are  aggregated coherently through propagation, which results in a stronger distortion compared to the NDM link. While the intrachannel signal--signal interaction is compensated for by the DBP algorithm, the signal--noise interaction remains. In Fig.~\ref{AIR_50G}\,(b), we remove the effects of SPM-induced NLPN by inserting all the ASE noise at the transmitter. It can be seen that the gap between the two AIRs  is closed. Also, an overall growth in the AIR is observed compared to Fig.~\ref{AIR_50G}\,(a), since the  effects of the signal--noise interaction are removed.

 As it is evident from Fig.~\ref{AIR_50G}\,(a), with AR(1),  higher AIRs can be obtained with the CDM link compared to the NDM link.
 To explain this, one should compare the spectral coherence of the XPM phase distortion depicted in Fig.~\ref{Fig_XPM_Corr_3ch}\,(d). As shown in the figure, with the CDM link the XPM spectral correlation is much higher than with the NDM link.  This strong frequency correlation indicates that XPM phase distortion $\theta(f,t)$ is independent of $f$ and can be modeled as a pure frequency-independent phase noise, such as  in \eqref{aux.ch}. Therefore, compared with the NDM link, with the CDM one, the XPM distortion can  handled more effectively by the detector optimized for the AR(1) model. 
  The AIR can be further improved by using the HOAR model, which accounts for the periodicity of the autocorrelation function in Fig.~\ref{Fig_XPM_Corr_3ch}\,(e).
  
   Fig.~\ref{AIR_28_100}\,(a) and (b) illustrate the results for $28$-GHz and $100$-GHz WDM grids. It can be seen that by increasing the WDM channel bandwidth, the gap between the NDM and  CDM links becomes more pronounced. This is because the  effects of the SPM-induced NLPN and the  frequency correlation of the XPM become stronger with increasing  bandwidth. 
 With a $100$-GHz WDM grid, should the AWGN auxiliary channel be used, the CDM link   is inferior to the NDM link by $4.5\%$ ($0.21$ bits) while with the AR auxiliary channels,  the CDM link surpasses the NDM link by $4.6\%$ ($0.23$ bits). Finally, Fig.~\ref{AIR_5ch} depicts the results for  five WDM channels. Comparing Fig.~\ref{AIR_5ch} to Fig.~\ref{AIR_50G}\,(a), we see that an increase in the number of channels has a negligible influence on the performance ranking across the three links. 

\section{Conclusions}\label{sec:conclusion}
We conducted a comparison between the performance of  CDM,  NDM, and DM links in terms of the AIR.  For the first time, we showed  that CDM links outperform NDM ones when a receiver that mitigates  XPM effects is deployed. This is  due to a higher XPM spectral coherence for CDM links. 
Moreover, our results indicate that, with a receiver optimized for an AWGN channel, which neglects the effects of XPM phase distortion,  CDM links are inferior to  NDM ones due to a higher SPM-induced NLPN.  
Finally, DM links were shown to be inferior to both  NDM and  CDM links, which is in accordance with the previous literature.

The results provided in this paper together with the  known advantages of  CDM links  in terms of  system complexity \cite{du2014channelized}, suggest that  CDM links implemented using FBGs, in combination with receivers that compensate for XPM, are promising candidates for a new generation of  WDM systems.  
Modern optical systems use polarization multiplexing  to transmit two complex signals at each WDM channel. Therefore, extending the results of this paper to polarization-multiplexed signals, which we leave to future studies, is of great practical interest.

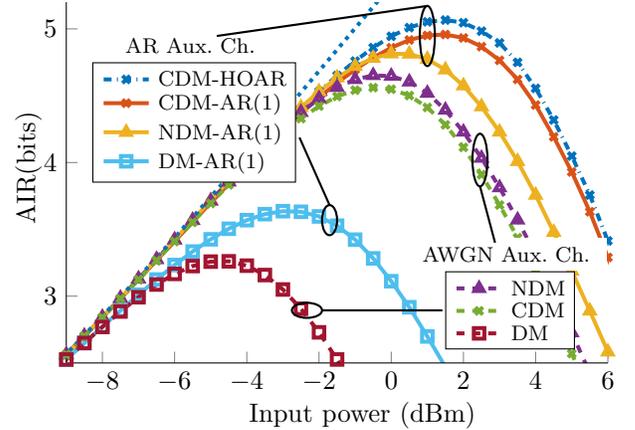
\begin{figure}[!t]
	\centering

	\begin{tikzpicture}[every text node part/.style={align=center}]

%
%
\definecolor{mycolor1}{rgb}{0.00000,0.44700,0.74100}%
\definecolor{mycolor2}{rgb}{0.85000,0.32500,0.09800}%
\definecolor{mycolor3}{rgb}{0.92900,0.69400,0.12500}%
\definecolor{mycolor4}{rgb}{0.49400,0.18400,0.55600}%
\definecolor{mycolor5}{rgb}{0.46600,0.67400,0.18800}%
\definecolor{mycolor6}{rgb}{0.30100,0.74500,0.93300}%
\definecolor{mycolor7}{rgb}{0.63500,0.07800,0.18400}%
\definecolor{mycolor8}{rgb}{0.2900,0.69400,0.12500}%

\begin{axis}[%
at={(0cm,0cm)},
enlargelimits=false,
axis on top,
width=9*.8cm,
height=6*.8cm,
scale only axis,
xmin=-9,
xmax=6,
ymin=2.5,
ymax=5.2,
axis background/.style={fill=white},
title style={font=\bfseries},
axis x line*=bottom,
axis y line*=left,
legend style={at={(0.047,.5)},font=\footnotesize,anchor=south west,legend cell align=left,align=left,draw=white!15!black}
]

\addplot [color=mycolor1,dashdotted,line width=1.5pt,mark=x,mark options={solid}]
table[row sep=crcr]{%
	-10	2.29455505848336\\
	-9.5	2.42739200124508\\
	-9	2.56304011419984\\
	-8.5	2.70052253220771\\
	-8	2.83905112195445\\
	-7.5	2.98121751834705\\
	-7	3.12422384014964\\
	-6.5	3.26837545381655\\
	-6	3.41245732008427\\
	-5.5	3.55772384250355\\
	-5	3.70217949941641\\
	-4.5	3.84613539403175\\
	-4	3.9884417250882\\
	-3.5	4.13145735382632\\
	-3	4.26891673213486\\
	-2.5	4.40289110189962\\
	-2	4.53164631756439\\
	-1.5	4.65077483288352\\
	-1	4.7589703327403\\
	-0.5	4.86023309628032\\
	0	4.94454411832405\\
	0.5	5.00960847348386\\
	1	5.05280728006335\\
	1.5	5.06627774107981\\
	2	5.04651067926136\\
	2.5	4.98309223655438\\
	3	4.89534839737374\\
	3.5	4.74954871943741\\
	4	4.55478722388028\\
	4.5	4.31852130681859\\
	5	4.04830789279214\\
	5.5	3.75681359137726\\
	6	3.41339262211661\\
	6.5	3.0525312431796\\
	7	2.66438593048144\\
	7.5	2.26704514181155\\
	8	1.85709797887709\\
	8.5	1.43183701026906\\
	9	1.05782606518564\\
	9.5	0.732186936683805\\
	10	0.444654609408274\\
};
\label{FBGAR32};

\addplot [color=mycolor2,solid,line width=1.5pt,mark=x,mark options={solid}]
table[row sep=crcr]{%
	-10	2.2951595069991\\
	-9.5	2.42779212755957\\
	-9	2.56304720947604\\
	-8.5	2.70062779105662\\
	-8	2.83996728494634\\
	-7.5	2.9812015864213\\
	-7	3.12363050583927\\
	-6.5	3.26734828076657\\
	-6	3.4115596159288\\
	-5.5	3.55548885647539\\
	-5	3.6988023004053\\
	-4.5	3.84068779459327\\
	-4	3.97976665020121\\
	-3.5	4.11472402727274\\
	-3	4.24388505078566\\
	-2.5	4.36785794313288\\
	-2	4.48643916334246\\
	-1.5	4.59620590068943\\
	-1	4.69587400785997\\
	-0.5	4.78444920685668\\
	0	4.85942571712977\\
	0.5	4.91706158009413\\
	1	4.95118762236733\\
	1.5	4.9580373976072\\
	2	4.93224764800674\\
	2.5	4.86712517988935\\
	3	4.76471421314773\\
	3.5	4.62215244473284\\
	4	4.43037144697821\\
	4.5	4.19214469555707\\
	5	3.92935880685165\\
	5.5	3.63151670670118\\
	6	3.28973060530691\\
	6.5	2.9377237177202\\
	7	2.56556943594433\\
	7.5	2.16341878317246\\
	8	1.76862075044287\\
	8.5	1.36120633995772\\
	9	0.995015621997487\\
	9.5	0.682960111252824\\
	10	0.425293158085934\\
};
\label{FBGAR1};

\addplot [color=mycolor3,solid,line width=1.5pt,mark=triangle,mark options={solid}]
table[row sep=crcr]{%
	-10	2.29226731403778\\
	-9.5	2.42578976132094\\
	-9	2.56209346871512\\
	-8.5	2.70091143913214\\
	-8	2.84198679083511\\
	-7.5	2.98501832590439\\
	-7	3.12967414775182\\
	-6.5	3.27561289987608\\
	-6	3.42247755337464\\
	-5.5	3.56974529480606\\
	-5	3.71671018334706\\
	-4.5	3.86297413825997\\
	-4	4.00710226307782\\
	-3.5	4.14797818904743\\
	-3	4.28371582098954\\
	-2.5	4.41224169878344\\
	-2	4.53041813298289\\
	-1.5	4.63445863583282\\
	-1	4.71974747565759\\
	-0.5	4.78131688492291\\
	0	4.81383519931976\\
	0.5	4.81235711022132\\
	1	4.77349684075667\\
	1.5	4.69398015192976\\
	2	4.57452052123743\\
	2.5	4.41792281159121\\
	3	4.22876779008835\\
	3.5	4.00826138391975\\
	4	3.75610889168917\\
	4.5	3.47835009639842\\
	5	3.18803880584493\\
	5.5	2.8895609322961\\
	6	2.58299208099788\\
	6.5	2.2744040806866\\
	7	1.95856267653758\\
	7.5	1.6445544807876\\
	8	1.33550352245173\\
	8.5	1.04074755800628\\
	9	0.756813788265664\\
	9.5	0.5049919977243\\
	10	0.299271576668442\\
};
\label{WDCAR1};

\addplot [color=mycolor4,dashed,line width=1.5pt,mark=triangle,mark options={solid}]
table[row sep=crcr]{%
	-10	2.29250675657555\\
	-9.5	2.42599379218553\\
	-9	2.56223694268721\\
	-8.5	2.70096842468204\\
	-8	2.84190075930375\\
	-7.5	2.98471819612758\\
	-7	3.12906376023197\\
	-6.5	3.27452050695233\\
	-6	3.42058513273544\\
	-5.5	3.56663156181231\\
	-5	3.7118615220621\\
	-4.5	3.85523851058421\\
	-4	3.9954011907692\\
	-3.5	4.1305529892145\\
	-3	4.25832872813739\\
	-2.5	4.37565142774326\\
	-2	4.47862007369685\\
	-1.5	4.56251313561634\\
	-1	4.62202326602899\\
	-0.5	4.65177304392182\\
	0	4.6469380307284\\
	0.5	4.60363612276158\\
	1	4.51917006725611\\
	1.5	4.39303313323922\\
	2	4.22867857792516\\
	2.5	4.03297308712798\\
	3	3.81225644886304\\
	3.5	3.57060272308215\\
	4	3.30947733401297\\
	4.5	3.02705532561592\\
	5	2.72684130891709\\
	5.5	2.42436123445655\\
	6	2.1310918861282\\
	6.5	1.8385705582335\\
	7	1.54745427713893\\
	7.5	1.26320315867351\\
	8	0.996811290606222\\
	8.5	0.74028733224655\\
	9	0.516406811757707\\
	9.5	0.325172898971702\\
	10	0.18025765669214\\
};
\label{Uncomp_AWGN};

\addplot [color=mycolor5,dashed,line width=1.5pt,mark=x,mark options={solid}]
table[row sep=crcr]{%
	-10	2.29546303257187\\
	-9.5	2.42806902925264\\
	-9	2.56326799367276\\
	-8.5	2.7007626910873\\
	-8	2.84023060046487\\
	-7.5	2.98131500138982\\
	-7	3.12361220562494\\
	-6.5	3.26665391470018\\
	-6	3.40988332478426\\
	-5.5	3.55262287959669\\
	-5	3.69402998723684\\
	-4.5	3.83303366574673\\
	-4	3.96823911427146\\
	-3.5	4.09778053328547\\
	-3	4.21910802116907\\
	-2.5	4.32874890001505\\
	-2	4.42223634242589\\
	-1.5	4.49458594774611\\
	-1	4.54150848638548\\
	-0.5	4.56038271590156\\
	0	4.54877076018592\\
	0.5	4.50126286056371\\
	1	4.41207939710731\\
	1.5	4.28193258441938\\
	2	4.11327645286021\\
	2.5	3.9087727771376\\
	3	3.6819091955942\\
	3.5	3.43751873395297\\
	4	3.16216604650381\\
	4.5	2.86497143131705\\
	5	2.57008875802525\\
	5.5	2.25850409264858\\
	6	1.94048225520685\\
	6.5	1.62525259552845\\
	7	1.33006423906957\\
	7.5	1.04040462964061\\
	8	0.765639737551157\\
	8.5	0.514449734191525\\
	9	0.324169779464383\\
	9.5	0.173152978475904\\
	10	0.0811001201139027\\
};
\label{FBG_AWGN};

\addplot [color=mycolor6,solid,line width=1.5pt,mark=square,mark options={solid}]
table[row sep=crcr]{%
	-10	2.27484180667501\\
	-9.5	2.39988467549502\\
	-9	2.52495705984725\\
	-8.5	2.64868125370851\\
	-8	2.77146874407193\\
	-7.5	2.89224735972269\\
	-7	3.01041851963903\\
	-6.5	3.12446065022042\\
	-6	3.23291372621113\\
	-5.5	3.33393736427681\\
	-5	3.4253160359244\\
	-4.5	3.5044770903748\\
	-4	3.56863821386389\\
	-3.5	3.61367111905179\\
	-3	3.63580785838424\\
	-2.5	3.63141313709384\\
	-2	3.5965798334624\\
	-1.5	3.52731209415178\\
	-1	3.42275698711854\\
	-0.5	3.28262744066727\\
	0	3.11177793997381\\
	0.5	2.91709005296627\\
	1	2.69835244101151\\
	1.5	2.46938521272385\\
	2	2.22914916035665\\
	2.5	1.98851933744094\\
	3	1.74602619539469\\
	3.5	1.51135259116667\\
	4	1.29117286562411\\
	4.5	1.08328199507402\\
	5	0.889163023572374\\
	5.5	0.712653252561323\\
	6	0.556159286201089\\
	6.5	0.420614178275036\\
	7	0.309955950667199\\
	7.5	0.219739402892855\\
	8	0.147444053005163\\
	8.5	0.0936539046264091\\
	9	0.0589154708100015\\
	9.5	0.0329432442999771\\
	10	0.0211678024191713\\
};
\label{DCF_AR1};

\addplot [color=mycolor7,dashed,line width=1.5pt,mark=square,mark options={solid}]
table[row sep=crcr]{%
	-10	2.27532434414926\\
	-9.5	2.40060863148676\\
	-9	2.52573195340978\\
	-8.5	2.64936956454281\\
	-8	2.76981548960373\\
	-7.5	2.88487839006084\\
	-7	2.99177315653759\\
	-6.5	3.08703158110292\\
	-6	3.1664733627437\\
	-5.5	3.22529700644442\\
	-5	3.25835605630502\\
	-4.5	3.26065891946187\\
	-4	3.22805455770655\\
	-3.5	3.15795778920082\\
	-3	3.04989274322573\\
	-2.5	2.90567080469394\\
	-2	2.72917580769575\\
	-1.5	2.52589248727781\\
	-1	2.30233981076926\\
	-0.5	2.06544763895057\\
	0	1.8219154965042\\
	0.5	1.57792652845544\\
	1	1.33940116904921\\
	1.5	1.11183460340668\\
	2	0.898940947175556\\
	2.5	0.702277485133002\\
	3	0.525302631725536\\
	3.5	0.374729178115055\\
	4	0.253403503761683\\
	4.5	0.158432919025664\\
	5	0.0894984261822171\\
	5.5	0.0449738908223391\\
	6	0.0195985459145123\\
	6.5	0.00739195446055027\\
	7	0.00240040030096648\\
	7.5	0.00062311907097289\\
	8	0.000279868976302308\\
	8.5	0.000163714170297905\\
	9	5.86083727386123e-05\\
	9.5	1.11014084536003e-05\\
	10	3.49326576883941e-06\\
};
\label{DCF_AWGN};

\addplot [color=mycolor1,dotted,line width=1.5pt]
table[row sep=crcr]{%
	-10	2.29756569921734\\
	-9.5	2.4313911296867\\
	-9	2.56810471639038\\
	-8.5	2.70749958236033\\
	-8	2.84937442894223\\
	-7.5	2.99353519187616\\
	-7	3.13979632014491\\
	-6.5	3.28798170676999\\
	-6	3.43792530801107\\
	-5.5	3.58947149129585\\
	-5	3.74247515330094\\
	-4.5	3.89680164857449\\
	-4	4.05232656655341\\
	-3.5	4.2089353913213\\
	-3	4.36652307442155\\
	-2.5	4.52499354682936\\
	-2	4.68425919204804\\
	-1.5	4.84424029840277\\
	-1	5.00486450506397\\
	-0.5	5.16606625319839\\
	0	5.3277862509359\\
	0.5	5.48997095854355\\
	1	5.65257209829042\\
	1.5	5.81554619192926\\
	2	5.97885412747266\\
	2.5	6.14246075596166\\
	3	6.30633451816985\\
	3.5	6.47044710062143\\
	4	6.63477311989082\\
	4.5	6.79928983386718\\
	5	6.96397687848127\\
	5.5	7.1288160282861\\
	6	7.29379097923602\\
	6.5	7.45888715200838\\
	7	7.62409151424498\\
	7.5	7.7893924201476\\
	8	7.95477946593571\\
	8.5	8.12024335975851\\
	9	8.28577580474375\\
	9.5	8.45136939395841\\
	10	8.61701751614875\\
};
\end{axis}
\node (Boxt) [draw,fill=white] at (5.9cm,.7cm) {\shortstack[l]{\footnotesize \ref{Uncomp_AWGN} NDM \\ \footnotesize \ref{FBG_AWGN}  CDM\\
		\footnotesize	\ref{DCF_AWGN}  DM
	}};
	\node [above=0cm of Boxt,fill=white]{\footnotesize AWGN Aux. Ch.  };
	\node (Boxt) [draw,fill=white] at (1.7cm,3.2cm) {\shortstack[l]{\footnotesize \ref{FBGAR32}  CDM-HOAR \\ \footnotesize \ref{FBGAR1}  CDM-AR(1) \\
			\footnotesize	\ref{WDCAR1} NDM-AR(1) \\
			\footnotesize	\ref{DCF_AR1} DM-AR(1)
		}};
		\node [above=0cm of Boxt,fill=white]{\footnotesize AR Aux. Ch.  };
		\draw[thick] (4.8,4.35) ellipse (.1cm and .4cm);
		\draw[thick] (2,4.35)--(4.8,4.75);
		\draw[thick] (5.5,2.7) ellipse (.1cm and .35cm);
		\draw[thick] (5.5,2.35)--(5.7,1.6);
		\draw[thick] (3.2,.7) ellipse (.2cm and .1cm);
		\draw[thick] (3.4,.7)--(5,.7);
		\draw[thick] (3.5,1.9) ellipse (.1cm and .2cm);
		\draw[thick] (3.5,2.1)--(3.1,3);
		\node at (4cm,-.7cm){Input power (dBm)  };
		\node[rotate=90] at (-.5cm,2.6cm){AIR(bits)};

	\end{tikzpicture}
	\caption{AIRs for a  50-GHz WDM grid with five copropagating channels.  }
	\label{AIR_5ch}
\end{figure}

%
%


\ifCLASSOPTIONcaptionsoff
  \newpage
\fi


\end{document}